# New Results on the Sum of Gamma Random Variates With Application to the Performance of Wireless Communication Systems over Nakagami-*m* Fading Channels

Imran Shafique Ansari, *Student Member, IEEE*, Ferkan Yilmaz, *Member, IEEE*, Mohamed-Slim Alouini, *Fellow, IEEE*, and Oğuz Kucur, *Member, IEEE*

*Abstract*—The probability density function (PDF) and cumulative distribution function of the sum of $L$ independent but not necessarily identically distributed Gamma variates, applicable to the output statistics of maximal ratio combining (MRC) receiver operating over Nakagami-$m$ fading channels or in other words to the statistical analysis of the scenario where the sum of squared Nakagami-$m$ distributions are user-of-interest, is presented in closed-form in terms of well-known Meijer's G function and easily computable Fox's $\bar{H}$ function for integer valued and non-integer valued $m$ fading parameters. Further analysis, particularly on bit error rate via a PDF-based approach is also offered in closed form in terms of Meijer's G function and Fox's $\bar{H}$ function for integer valued fading parameters, and extended Fox's $\bar{H}$ function ($\hat{H}$) for non-integer valued fading parameters. Our proposed results complement previous known results that are either expressed in terms of infinite sums, nested sums, or higher order derivatives of the fading parameter $m$.

## I. INTRODUCTION

IN recent times, different diversity schemes have marked an important impact in the arena of wireless communication systems. The main reason behind this is that these different diversity schemes allow for multiple transmission and/or reception paths for the same signal [1]. One of the optimal diversity combining scheme is the maximal ratio combining (MRC) diversity scheme where all the diversity branches are processed to obtain the best possible devised and improved single output that possibly stays above a certain specified threshold [1]–[3].

Additionally, wireless communications are driven by a complicated phenomenon known as radio-wave propagation that is characterized by various effects such as fading, shadowing and path-loss. The statistical behavior of these effects is described by different models depending on the nature of the communication environment. The wide versatility, experimental validity, and analytical tractability of Nakagami-$m$ distribution [4] has made it a very popular fading model for performance analysis investigations in diversity schemes of wireless communications (for instance, [5] among others and [6] and references therein). In addition, it is useful to mention that Nakagami-$m$ distribution is useful to study multihop relay networks [7], [8]. Hence, this and such other distributions have many other applications in wireless communication engineering problems and one of those that we focus on is a communication system employing MRC diversity scheme undergoing this distribution i.e. the study of MRC diversity combining receiver operating over Nakagami-$m$ fading channels [4], where the statistics of the sum of Gamma random variates (RVs) or equivalently the sum of squared Nakagami-$m$ RVs are required and moreover, the performance analysis of such wireless communication systems usually requires complicated and tedious tasks related to statistics as elegantly explained in details in [9].

The probability density function (PDF) and cumulative distribution function (CDF) of the sum of $L$ independent but not necessarily identical (i.n.i.d.) Gamma RVs have been investigated quite extensively in the past in [6], [9]–[11] but the published results are sometimes given in rather complicated expressions in the form of single definite or indefinite series that renders the given expressions therein not always computationally efficient. In more details, Moschopoulos has proposed in [11] an infinite-series representation for the PDF of the sum of the i.n.i.d. Gamma RVs and Alouini *et. al* have extended the results of [11] in [6] for the case of arbitrarily correlated Gamma RVs and studied the performance of MRC among other receivers. Commonly, the moment generating function (MGF)-based approach or characteristic function (CF)-based approach have been followed for the performance analysis and derived accurate and/or approximate analytical results in terms of either infinite sums and/or higher order derivatives of the diversity order [12]–[17]. This occurs as there are no simple closed-form expressions available in the open literature either for the PDF or the CDF of the sum of i.n.i.d. Gamma RVs [9]. In [18], the authors have also introduced the statistics of sum of multiple Gamma RVs with correlation but have considered only the case wherein the RVs are identically distributed. Recently, Karagiannidis *et. al* have obtained in

Imran Shafique Ansari, Ferkan Yilmaz and Mohamed-Slim Alouini are with King Abdullah University of Science and Technology (KAUST), Al-Khawarizmi Applied Math. Building (Bldg. #1), Thuwal, Makkah Province, Kingdom of Saudi Arabia (e-mail: {imran.ansari, ferkan.yilmaz, slim.alouini}@kaust.edu.sa).

Oğuz Kucur is with Gebze Institute of Technology (GYTE), P.K 141, 41400, Gebze, Kocaeli, Turkey (e-mail: okucur@gyte.edu.tr).

[9] closed-form expressions for the PDF and the CDF of the sum of nonidentical squared Nakagami-*m* RVs or equivalently Gamma RVs with integer-order fading parameters but these results involve a series of nested summations that can be computationally complex and expensive.

In this work, we offer novel closed-form expressions for the PDF and CDF of the sum of i.n.i.d. Gamma RVs or equivalently squared Nakagami-*m* RVs with integer-order as well as non-integer-order fading parameters in terms of easily computable Meijer's G function [19] and Fox's H̄ function [20]–[22, App. (A.5)], respectively. It is noteworthy to mention that the bit error rate (BER) is one of the most important performance measures that forms the basis in designing wireless communication systems. Hence, we demonstrate closed-form expressions of the BER, as a performance metric, for binary modulation schemes, via a PDF-based approach, of a $L$-branch MRC diversity receiver in the presence of Gamma or Nakagami-*m* multipath fading, in terms of Meijer's G function and Fox's H̄ function for integer-order fading parameters, and in terms of extended Fox's H̄ function ($\hat{H}$) [23][1] for non-integer-order fading parameters. This proves the importance and the simplicity in the employment of those earlier derived simple closed-form statistical PDF and CDF expressions. These resulting expressions also give alternative closed-forms for previously known/published results obtained via CF or MGF-based approaches. It should be noted that all our newly proposed results are readily computable by Mellin-Barnes theorem that further corroborates the generality and the usefulness of the analytical frameworks introduced in this paper. Finally, it must be further mentioned that these results have been checked and validated by Monte Carlo simulations.

The remainder of the paper is organized as follows. Section II introduces the system and Section III gives novel closed-form expressions for the PDF and CDF of the sum of Gamma or equivalently squared Nakagami-*m* RVs in terms of Meijer's G function and Fox's H̄ function respectively for integer-order and non-integer-order fading parameters respectively. Next, Section IV utilizes these results presented in Section III to derive useful expressions for the BER, as a performance metric, for MRC diversity receivers operating over i.n.i.d. Gamma fading channels or equivalently Nakagami-*m* diversity paths in terms of Meijer's G function and Fox's H̄ function for integer-order fading parameters, and extended Fox's H̄ function ($\hat{H}$) for non-integer-order fading parameters. Further, Section IV also discusses the results followed by the summary of the paper in the last section.

## II. NAKAGAMI-*m* CHANNEL MODEL AND GAMMA DISTRIBUTION

A MRC based communication system with a source and a destination is considered with $L$ diversity paths undergoing i.n.i.d. Nakagami-*m* fading channels as follows

$$Y_l = \alpha_l X + n_l, \qquad l = 1, 2, \ldots, L, \qquad (1)$$

[1]The extended Fox's H̄ function ($\hat{H}$), the Fox's H̄ function, the Fox's H function, and the Meijer's G function are extensively defined in Table I.

where $Y_l$ is the received signal at the $l$-th branch receiver end, $X$ is the transmitted signal, $\alpha_l$ is the channel gain, and $n_l$ is the additive white Gaussian noise (AWGN). In a Nakagami-*m* multipath fading channel, $\gamma_l = |\alpha_l|^2$ follows a Gamma distribution. Hence, the channel gains experience multipath fading whose statistics follows a Gamma distribution with PDF given by

$$p_{\gamma_l}(\gamma) = \left(\frac{m_l}{\Omega_l}\right)^{m_l} \frac{\gamma_l^{m_l-1}}{\Gamma(m_l)} e^{-\frac{m_l}{\Omega_l}\gamma_l}, \qquad (2)$$

where $m_l > 0$ and $\Omega_l > 0$ are known as fading figure representing the diversity order of the fading environment and the mean of the local power, respectively, and where $\Gamma(\cdot)$ denotes the Gamma function [24, Eq. (8.310)]. In more details, the parameter $m_l$ quantifies the severity of multipath fading, in the sense that small values of $m$ indicates severe multipath fading and vice versa. The instantaneous signal-to-noise ratio (SNR) of the $l$th branch is given by $\gamma_l = (E_b/N_0)|\alpha_l|^2$, $E_b$ is the average energy per bit, and $N_0$ is the one sided power spectral density of the AWGN.

## III. CLOSED-FORM STATISTICAL CHARACTERISTICS FOR THE SUM OF GAMMA RANDOM VARIATES

This section presents the results on the statistical characteristics including the PDF and CDF of the sum of i.n.i.d. Gamma variates. To best of the author's knowledge, it is useful to mention again that the PDF of the sum of Gamma distributions given in the following theorem is a novel closed-form result not reported in the literature earlier. It includes special cases that are used in the literature such as independent and identically distributed (i.i.d.) Gamma RVs and/or integer-fading figure parameters among others [6], [9].

### A. PDF

*1) General Case (Non-Integer m Fading Parameters):*

**Theorem 1** (PDF of the Sum of Gamma or Equivalently Squared Nakagami-*m* RVs). *Let $\{\gamma_l\}_{l=1}^{L}$ be a set of i.n.i.d. Gamma variates with parameters $m_l$ and $\Omega_l$[2]. Then, the closed-form PDF of the sum*

$$Y = \sum_{l=1}^{L} \gamma_l \qquad (3)$$

*for both integer-order as well as non-integer-order fading*

[2]For correlated diversity branches, the statistical characteristics derivation and the performance analysis can be carried out in a similar fashion as in the independent fading case. For an arbitrarily correlated Nakagami-*m* fading environment, assuming that the fading parameter is common to all the diversity branches, the desired MGF of the sum of correlated Gamma RVs can be expressed as $\mathcal{M}_Y(s) = \det(I + sR\Lambda)^{-m} = \prod_{l=1}^{L} \mathcal{M}_{\gamma_l}(s) = \prod_{l=1}^{L}(1 + \lambda_l s)^{-m}$ where $I$ is the $L \times L$ identity matrix, $\Lambda$ is a positive definite matrix of dimension $L$ (determined by the branch covariance matrix), $R$ is a diagonal matrix as $R = \text{diag}(\Omega_1/m, \ldots, \Omega_L/m)$, and $\lambda_l$ is the $l$th eigenvalue of matrix $R\Lambda$ where each eigenvalue is modeled as a Gamma RV. Hence, on replacing $\gamma_l$'s with $\lambda_l$'s, in our presented work, we will achieve the results applicable to the correlated diversity case.

TABLE I
REPRESENTATION OF THE EXTENDED FOX'S $\bar{H}$ FUNCTION ($\hat{H}$) AND ITS SPECIAL CASES

The extended Fox's $\bar{H}$ function ($\hat{H}$) is defined by [23] as

$$\hat{H}_{p,q}^{m,n}\left[z \left| \begin{array}{c} (\alpha_j, A_j, a_j)_{1,p} \\ (\beta_j, B_j, b_j)_{1,q} \end{array} \right. \right] = \frac{1}{2\pi i} \oint_C \frac{\prod_{j=1}^{n}\{\Gamma(1-\alpha_j+A_js)\}^{a_j} \prod_{j=1}^{m}\{\Gamma(\beta_j-B_js)\}^{b_j}}{\prod_{j=n+1}^{p}\{\Gamma(\alpha_j-A_js)\}^{a_j} \prod_{j=m+1}^{q}\{\Gamma(1-\beta_j+B_js)\}^{b_j}} z^s ds, \quad \text{(T.I.1)}$$

which contains fractional powers of $\Gamma$-functions. Here $z$ may be real or complex but is not equal to zero and an empty product is interpreted as unity; $C$ is a suitable contour, and positive integers $p$, $q$, $m$, and $n$ satisfy the following inequalities: $1 \leq m \leq q$, $0 \leq n \leq p$, $A_j > 0 (j=1,\ldots,p)$, $B_j > 0 (j=1,\ldots,q)$ and $\alpha_j (j=1,\ldots,p)$, and $\beta_j (j=1,\ldots,q)$ are complex parameters. The exponents $a_j (j=1,\ldots,p)$ and $b_j (j=1,\ldots,q)$ can take on non-integer values. The poles of this integrand are assumed to be simple and the contour in this definition is presumed to be imaginary axis Re($s$)=0 that is suitably intended in order to avoid the singularities of the Gamma functions and to keep these singularities at appropriate sides.

When the exponents $a_j = 1$ for $(j=n+1,\ldots,p)$ and $b_j = 1$ for $(j=1,\ldots,m)$, the extended Fox's $\bar{H}$ function ($\hat{H}$) reduces to the familiar Fox's $\bar{H}$ function defined by [20]–[22] as

$$\bar{H}_{p,q}^{m,n}\left[z \left| \begin{array}{c} (\alpha_j, A_j, a_j)_{1,n}, (\alpha_j, A_j)_{n+1,p} \\ (\beta_j, B_j)_{1,m}, (\beta_j, B_j, b_j)_{m+1,q} \end{array} \right. \right] = \frac{1}{2\pi i} \oint_C \frac{\prod_{j=1}^{n}\{\Gamma(1-\alpha_j+A_js)\}^{a_j} \prod_{j=1}^{m}\Gamma(\beta_j-B_js)}{\prod_{j=n+1}^{p}\Gamma(\alpha_j-A_js) \prod_{j=m+1}^{q}\{\Gamma(1-\beta_j+B_js)\}^{b_j}} z^s ds, \quad \text{(T.I.2)}$$

which contains fractional powers of $\Gamma$-functions. Here $z$ may be real or complex but is not equal to zero and an empty product is interpreted as unity; $C$ is a suitable contour, and positive integers $p$, $q$, $m$, and $n$ satisfy the following inequalities: $1 \leq m \leq q$, $0 \leq n \leq p$, $A_j > 0 (j=1,\ldots,p)$, $B_j > 0 (j=1,\ldots,q)$ and $\alpha_j (j=1,\ldots,p)$, and $\beta_j (j=1,\ldots,q)$ are complex parameters. The exponents $a_j (j=1,\ldots,n)$ and $b_j (j=m+1,\ldots,q)$ can take on non-integer values. The poles of this integrand are assumed to be simple and the contour in this definition is presumed to be imaginary axis Re($s$)=0 that is suitably intended in order to avoid the singularities of the Gamma functions and to keep these singularities at appropriate sides.

Now, when the exponents $a_j = b_j = 1 \forall j$, the Fox's $\bar{H}$ function reduces to the familiar Fox's H-function defined by [19] as

$$H_{p,q}^{m,n}\left[z \left| \begin{array}{c} (\alpha_1, A_1),\ldots,(\alpha_p, A_p) \\ (\beta_1, B_1),\ldots,(\beta_q, B_q) \end{array} \right. \right] = \frac{1}{2\pi i} \oint_C \frac{\prod_{j=1}^{n}\Gamma(1-\alpha_j+A_js) \prod_{j=1}^{m}\Gamma(\beta_j-B_js)}{\prod_{j=n+1}^{p}\Gamma(\alpha_j-A_js) \prod_{j=m+1}^{q}\Gamma(1-\beta_j+B_js)} z^s ds, \quad \text{(T.I.3)}$$

which contains fractional powers of $\Gamma$-functions. Here $z$ may be real or complex but is not equal to zero and an empty product is interpreted as unity; $C$ is a suitable contour, and positive integers $p$, $q$, $m$, and $n$ satisfy the following inequalities: $1 \leq m \leq q$, $0 \leq n \leq p$, $A_j > 0 (j=1,\ldots,p)$, $B_j > 0 (j=1,\ldots,q)$ and $\alpha_j (j=1,\ldots,p)$, and $\beta_j (j=1,\ldots,q)$ are complex parameters. The poles of this integrand are assumed to be simple and the contour in this definition is presumed to be imaginary axis Re($s$)=0 that is suitably intended in order to avoid the singularities of the Gamma functions and to keep these singularities at appropriate sides.

Finally, when the exponents $A_j = B_j = 1 \forall j$, the Fox's $\bar{H}$ function reduces to the familiar Meijer's G function defined by [19] as

$$G_{p,q}^{m,n}\left[z \left| \begin{array}{c} \alpha_1,\ldots,\alpha_p \\ \beta_1,\ldots,\beta_q \end{array} \right. \right] = \frac{1}{2\pi i} \oint_C \frac{\prod_{j=1}^{n}\Gamma(1-\alpha_j+s) \prod_{j=1}^{m}\Gamma(\beta_j-s)}{\prod_{j=n+1}^{p}\Gamma(\alpha_j-s) \prod_{j=m+1}^{q}\Gamma(1-\beta_j+s)} z^s ds, \quad \text{(T.I.4)}$$

which contains fractional powers of $\Gamma$-functions. Here $z$ may be real or complex but is not equal to zero and an empty product is interpreted as unity; $C$ is a suitable contour, and positive integers $p$, $q$, $m$, and $n$ satisfy the following inequalities: $1 \leq m \leq q$, and $0 \leq n \leq p$. $\alpha_j (j=1,\ldots,p)$, and $\beta_j (j=1,\ldots,q)$ are complex parameters. The poles of this integrand are assumed to be simple and the contour in this definition is presumed to be imaginary axis Re($s$)=0 that is suitably intended in order to avoid the singularities of the Gamma functions and to keep these singularities at appropriate sides.

---

parameters can be expressed in terms of the Fox's $\bar{H}$ function[3] as

$$p_Y(y) = \prod_{l=1}^{L} \left(\frac{m_l}{\Omega_l}\right)^{m_l} \bar{H}_{L,L}^{0,L}\left[e^y \left| \begin{array}{c} \Xi_L^{(1)} \\ \Xi_L^{(2)} \end{array} \right. \right], \quad (4)$$

where $y > 0$, the coefficient sets $\Xi_k^{(1)}$ and $\Xi_k^{(2)}$, $k \in \mathbb{N}$ are defined as

$$\Xi_k^{(1)} = \overbrace{\left(1-\frac{m_1}{\Omega_1}, 1, m_1\right), \ldots, \left(1-\frac{m_k}{\Omega_k}, 1, m_k\right)}^{k\text{-bracketed terms}}, \quad (5)$$

and

$$\Xi_k^{(2)} = \overbrace{\left(-\frac{m_1}{\Omega_1}, 1, m_1\right), \ldots, \left(-\frac{m_k}{\Omega_k}, 1, m_k\right)}^{k\text{-bracketed terms}}, \quad (6)$$

[3]To our best knowledge, the Fox's $\bar{H}$ function [20]–[22, App. (A.5)] is not available in any standard mathematical packages. As such, we offer in Table II an efficient MATHEMATICA® implementation of this function (similar to [7], [25], [26]) in order to give numerical results based on (4). With this implementation, the Fox's $\bar{H}$ function can be evaluated fast and accurately. This computability, therefore, has been utilized for different scenarios and is employed to discuss the results in comparison to respective Monte Carlo simulation outcomes.

respectively.

*Proof:* In order to derive the PDF of $Y$, we proceed as follows. Firstly, the MGF

$$\mathcal{M}_{\gamma_l}(s) \triangleq \mathbb{E}\left[e^{-\gamma_l s}\right] = \int_0^\infty e^{-\gamma_l s} p_{\gamma_l}(\gamma) d\gamma \quad (7)$$

of a single Gamma distribution is given as [1, Eq. (2.22)]

$$\mathcal{M}_{\gamma_l}(s) = \left(1 + \frac{\Omega_l}{m_l}s\right)^{-m_l}. \quad (8)$$

Then, after performing some simple algebraic manipulations using [27, Eq. (6.1.15)], we can rewrite the MGF of a single Gamma distribution as

$$\mathcal{M}_{\gamma_l}(s) = \left(\frac{m_l}{\Omega_l}\right)^{m_l} \frac{\Gamma^{m_l}\left(\frac{m_l}{\Omega_l}+s\right)}{\Gamma^{m_l}\left(1+\frac{m_l}{\Omega_l}+s\right)}. \quad (9)$$

Since, $\gamma_l's$ are independent, the MGF of Y is the product of the MGF's of the $\gamma_l's$, i.e.

$$\mathcal{M}_Y(s) = \prod_{l=1}^{L} \mathcal{M}_{\gamma_l}(s). \quad (10)$$

Now, we express the PDF of the sum of Gamma RVs, using the obtained MGF in (10), via the inverse Laplace transform [28]

$$p_Y(y) = \mathcal{L}^{-1}\{\mathcal{M}(s)\} = \frac{1}{2\pi i}\oint_C \mathcal{M}_Y(s)e^s ds \quad (11)$$

that produces a Mellin-Barnes contour integral [29] representation as (similar to (T.I.1), Table I)

$$p_Y(y) = \prod_{l=1}^{L}\left(\frac{m_l}{\Omega_l}\right)^{m_l} \times \frac{1}{2\pi i}\oint_C \frac{\prod_{l=1}^{L}\Gamma^{m_l}\left(\frac{m_l}{\Omega_l}+s\right)}{\prod_{l=1}^{L}\Gamma^{m_l}\left(1+\frac{m_l}{\Omega_l}+s\right)}e^s ds. \quad (12)$$

Hence we use this obtained result and perform some simple rearrangements on the $\Gamma(.)$ terms in the Mellin-Barnes contour integral representation to express the closed-form PDF of the sum of Gamma RVs $Y$, valid for both integer-order as well as non-integer-order fading parameters, in terms of Fox's $\bar{H}$ function as given in (4). ∎

It is worthy mentioning that the PDF of sum of Gamma or equivalently squared Nakagami-$m$ RVs with arbitrary fading parameters has also been successfully achieved in terms of the confluent form of the multivariate Lauricella hypergeometric function and given by [14, Eq. (35)]

$$p_Y(y) = \frac{1}{\Gamma\left(\sum_{l=1}^{L}m_l\right)}\left[\prod_{l=1}^{L}\left(\frac{m_l}{\Omega_l}\right)^{m_l}\right]y^{\left(\sum_{l=1}^{L}m_l\right)-1}$$
$$\times \phi_2^{(L)}\left(m_1,\ldots,m_L;\sum_{l=1}^{L}m_l;-\frac{m_1}{\Omega_1}y,\ldots,-\frac{m_l}{\Omega_l}y\right), \quad (13)$$

where the confluent Lauricella hypergeometric function $\phi_2^{(n)}(\ldots)$ is defined as [14, Eq. (36)], [30]

$$\phi_2^{(n)}(b_1,\ldots,b_n;c;x_1,\ldots,x_n)$$
$$= \underbrace{\sum_{i_1=0}^{\infty}\cdots\sum_{i_n=0}^{\infty}}_{n\text{-fold summation}}\frac{(b_1)_{i_1}\ldots(b_n)_{i_n}}{(c)_{i_1+\cdots+i_n}}\frac{x_1^{i_1}}{i_1!}\cdots\frac{x_n^{i_n}}{i_n!}, \quad (14)$$

and involves as such an $L$-fold infinite summations. On the other hand, our alternative result presented in Theorem 1 involves only one single-fold integration.

The motivation and the possibility that led to the above result presented in Theorem 1 was the representation of the PDF a single Gamma RV in terms of Meijer's G function as discussed below in Corollary 1.

**Corollary 1** (PDF of a Single Gamma RV). *Let $\{\gamma_l\}$ be any i.n.i.d. Gamma variate with parameters $m_l$ and $\Omega_l$. Then, the closed-form PDF of this single Gamma RV for integer-order fading parameters can be expressed as*

$$p_{\gamma_l}(\gamma) = \left(\frac{m_l}{\Omega_l}\right)^{m_l}G_{m_l,m_l}^{m_l,0}\left[e^{-\gamma_l}\left|\begin{array}{c}\Phi_{m_l}^{(1)}\\ \Phi_{m_l}^{(2)}\end{array}\right.\right], \quad (15)$$

*where the coefficient sets $\Phi_k^{(1)}$ and $\Phi_k^{(2)}$, $k\in\mathbb{N}$ are defined as*

$$\Phi_k^{(1)} = \overbrace{\left(1+\frac{m_k}{\Omega_k}\right),\ldots,\left(1+\frac{m_k}{\Omega_k}\right)}^{k\text{-times}}, \quad (16)$$

*and*

$$\Phi_k^{(2)} = \overbrace{\left(\frac{m_k}{\Omega_k}\right),\ldots,\left(\frac{m_k}{\Omega_k}\right)}^{k\text{-times}}, \quad (17)$$

*respectively.*

*Proof:* The PDF of a single Gamma RV for integer-order fading parameters can be expressed by placing $L=1$ in (4). This substitution gets simplified, via simple algebraic manipulations, to the above obtained result in (15). Alternatively, the PDF of a single Gamma RV for integer-order fading parameters can be expressed using the obtained MGF in (9), via inverse Laplace transform [28] that produces a Mellin-Barnes integral [29] representation. Hence we use this obtained result to express the PDF of a single Gamma RV in an alternative form, in terms of Meijer's G function, as expressed in (15). ∎

TABLE II
MATHEMATICA® IMPLEMENTATION OF THE FOX'S $\bar{H}$ FUNCTION

```
(*Fox H-Bar-Function Implementation*)
Clear[x, Ω];
(*Exception*)
FoxHBar::InconsistentCoeffs = "Inconsistent coefficients!";
FoxHBar[a_, b_, z_] := Module[
    {Z, s, Pa, Pb, Qa, Qb, M, R, value},
    (*Gamma product terms*)
    Pa = Function[u, Product[
        Power[Gamma[1 - a[[1, n, 1]] + u a[[1, n, 2]]], a[[1, n, 3]]], {n, 1, Length[a[[1]]]}]];
    Qa = Function[u, Product[Gamma[a[[2, n, 1]] - u a[[2, n, 2]]], {n, 1, Length[a[[2]]]}]];
    Pb = Function[u, Product[Gamma[b[[1, n, 1]] - u b[[1, n, 2]]], {n, 1, Length[b[[1]]]}]];
    Qb = Function[u, Product[
        Power[Gamma[1 - b[[2, n, 1]] + u b[[2, n, 2]]], b[[2, n, 3]]], {n, 1, Length[b[[2]]]}]];
    M = Function[u, Pa[u] Pb[u] / Qa[u] / Qb[u]];
    (*Contour Limiter*)
    (*Depends on numerator argument i.e. it
     must be at least half of the least valued gamma arguments*)
    R = -1;
    (*Assignment and Declaration*)
    Z = z;
    (*Final Evaluation*)
    value = 1/(2 π I)
      NIntegrate[M[s] Z^s, {s, -50 - I 100, R - I 100, R + I 100, -50 + I 100}, MaxRecursion → 55];
    (*Returning back the value*)
    Return[value];
];
(*End of FoxHBar*)
```

As for validation and numerical examples, Fig. 1 and Fig. 2 present the PDF and the logarithmic PDF respectively of the output SNR obtained from the exact closed-form expression (4) and show a perfect match between this obtained closed-form analytical result and the one obtained via Monte Carlo simulations for varying $L's$ (i.e. $L=3,4,5$), and their respective fixed fading parameters $m_1=0.6$, $m_2=1.1$, $m_3=2$, $m_4=3.4$, and $m_5=4.5$.

For additional verification purposes, we simplified the PDF expression presented in [9, Theorem 1, Eq. (6)] for the sum of squared Nakagami-$m$ or equivalently Gamma RVs, for a special case with $L = 2$ i.e. a dual-branch MRC diversity combining receiver based wireless communication system undergoing Rayleigh fading i.e. $m_1 = 1$ and $m_2 = 1$. We obtained the following simplified PDF expression.

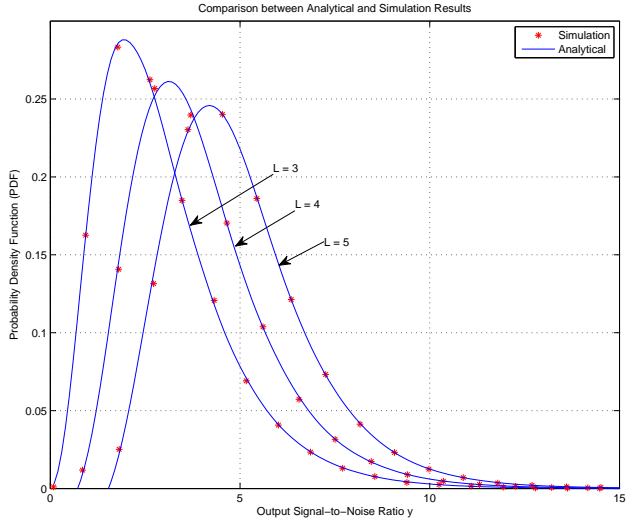

Fig. 1. Comparison between PDFs obtained analytically and via Monte Carlo simulations for varying branches $L$ and respective fixed fading parameters for these channels with $m_1 = 0.6$, $m_2 = 1.1$, $m_3 = 2$, $m_4 = 3.4$, and $m_5 = 4.5$.

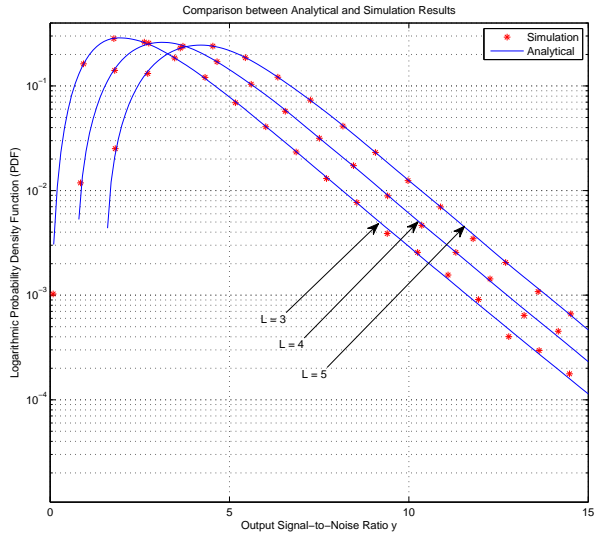

Fig. 2. Comparison between PDFs obtained analytically and via Monte Carlo simulations, on log scale, for varying branches $L$ and respective fixed fading parameters for these channels with $m_1 = 0.6$, $m_2 = 1.1$, $m_3 = 2$, $m_4 = 3.4$, and $m_5 = 4.5$.

$$f_{Z_L}(z) = \frac{1}{\eta_1 - \eta_2} \left[ e^{-\frac{z}{\eta_1}} - e^{-\frac{z}{\eta_2}} \right], \quad (18)$$

where, $\eta'_l$s are equivalent to our $\Omega'_l$s and $z$ is the RV as opposed to our RV $y$. This obtained expression, when plotted against our results, with similar specific values, found a perfect match along with the Monte Carlo simulations and hence further conforming our results as shown in Fig. 3.

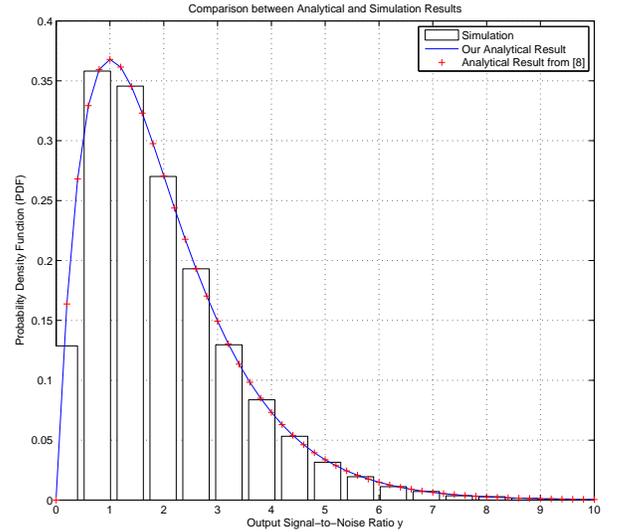

Fig. 3. Comparison between PDFs obtained analytically, via Monte Carlo simulations, and via [9, Theorem 1, Eq. (6)] with $L = 2$ and fading parameters for these channels with $m_1 = 1$ and $m_2 = 1$ i.e. Rayleigh fading channels.

*2) Special Case (Integer $m$ Fading Parameters):*

**Corollary 2** (PDF of the Sum of Gamma or Equivalently Squared Nakagami-$m$ RVs for Integer-Order Fading Parameters). *It is worth mentioning that the closed-form expression in* (4) *simplifies to the following expression* (20) *for integer-order fading parameters via simple algebraic manipulations.*

$$p_Y(y) = \prod_{l=1}^{L} \left( \frac{m_l}{\Omega_l} \right)^{m_l} G_{\kappa,\kappa}^{\kappa,0} \left[ e^{-y} \left| \begin{array}{c} \Psi_\kappa^{(1)} \\ \Psi_\kappa^{(2)} \end{array} \right. \right], \quad (20)$$

*where,*

$$\kappa = \sum_{l=1}^{L} m_l \quad (21)$$

*is an integer, the coefficient set $\Psi_k^{(1)}$, $k \in \mathbb{N}$ is as defined in* (19) *and the coefficient set $\Psi_k^{(2)}$, $k \in \mathbb{N}$ is defined as*

$$\Psi_k^{(2)} = \Big( \overbrace{\underbrace{\left(\frac{m_1}{\Omega_1}\right), \ldots, \left(\frac{m_1}{\Omega_1}\right)}_{m_1\text{-times}}, \ldots, \underbrace{\left(\frac{m_K}{\Omega_K}\right), \ldots, \left(\frac{m_K}{\Omega_K}\right)}_{m_K\text{-times}}}^{k\text{-bracketed terms}} \Big), \quad (22)$$

*where $K$ is the total number of Gamma or equivalently squared Nakagami-$m$ RVs i.e. $L$ number of total branches for our specific wireless communication system being considered here.*

At this point it should be mentioned that the PDF of the sum of Gamma or equivalently squared Nakagami-$m$ RVs for integer-order fading parameters was also proposed by Karagiannidis *et. al.* in [9, Eq. (6)]. More precisely, our result

$$\Psi_k^{(1)} = \overbrace{\overbrace{\left(1+\frac{m_1}{\Omega_1}\right), \ldots, \left(1+\frac{m_1}{\Omega_1}\right)}^{m_1\text{-times}}, \ldots, \overbrace{\left(1+\frac{m_K}{\Omega_K}\right), \ldots, \left(1+\frac{m_K}{\Omega_K}\right)}^{m_K\text{-times}}}^{k\text{-bracketed terms}} \quad (19)$$

in Corollary 2 can also be represented in terms of a series of $L-2$ nested weighted summations of Erlang PDFs as per [9, Eq. (6)]. On the other hand, the result presented in Corollary 2 offers an alternative representation that involves only one single-fold integration (in terms of Meijer's G function that is readily available in the standard mathematical packages such as MATHEMATICA, MATLAB and MAPLE).

Now let us consider some special cases in order to check the correctness and accuracy of (20). These special cases give a further insight to the above obtained results and assist in understanding the rest of the results presented in this work.

**Special Case 1** (Sum of Two Exponential RVs). Let us assume that we have two i.n.i.d. Gamma RVs with fading figures $m_1 = 1$ and $m_2 = 1$ and average powers $\Omega_1$ and $\Omega_2$. Substituting these parameters in (20) results in

$$p_Y(y) = \frac{1}{\Omega_1 \Omega_2} \mathrm{G}_{2,2}^{2,0}\left[e^{-y} \left| \begin{array}{c} 1+\frac{1}{\Omega_1}, 1+\frac{1}{\Omega_2} \\ \frac{1}{\Omega_1}, \frac{1}{\Omega_2} \end{array} \right. \right]. \quad (23)$$

Then, using the Meijer's G identity given in [22, Eq. (1.142)] and then using [31, Eq. (07.23.03.0227.01)], (23) readily reduces to [32, Sec. 5.2.4]

$$p_Y(y) = \frac{e^{-\frac{y}{\Omega_1}} - e^{-\frac{y}{\Omega_2}}}{\Omega_1 - \Omega_2}, \quad (24)$$

as expected.

**Special Case 2** (Sum of $L$ Exponential RVs). Let us assume that we have $L$ i.n.i.d. Gamma RVs with fading figures $m_l = 1$ for all $l \in \{1,2,3,\ldots,L\}$ and average powers $\Omega_l \neq \Omega_k$ for all $k,l \in \{1,2,3,\ldots,L\}$. Substituting these parameters in (20) results in

$$p_Y(y) = \frac{1}{\prod_{l=1}^L \Omega_l} \mathrm{G}_{L,L}^{L,0}\left[e^{-y} \left| \begin{array}{c} 1+\frac{1}{\Omega_1}, \ldots, 1+\frac{1}{\Omega_L} \\ \frac{1}{\Omega_1}, \ldots, \frac{1}{\Omega_L} \end{array} \right. \right]. \quad (25)$$

Then, performing some algebraic manipulations using the Meijer's G identity given in [31, Eq. (07.34.26.0004.01)] and [31, 07.31.06.0017.01], we simplify (25) to [32, Eq. (5.8)]

$$p_Y(y) = \sum_{l=1}^L \left( \prod_{k\neq l} \frac{\frac{1}{\Omega_k}}{\frac{1}{\Omega_k} - \frac{1}{\Omega_l}} \right) \frac{1}{\Omega_l} e^{-\frac{y}{\Omega_l}}, \quad (26)$$

as expected.

### B. CDF

*1) General Case (Non-Integer m Fading Parameters):*

**Theorem 2** (CDF of the Sum of Gamma or Equivalently Squared Nakagami-m RVs). *The CDF of $Y$ for both integer-order as well as non-integer-order fading parameters can be closely expressed in terms of the Fox's $\bar{H}$ function as*

$$P_Y(y) = 1 + \prod_{l=1}^L \left(\frac{m_l}{\Omega_l}\right)^{m_l} \bar{\mathrm{H}}_{L+1,L+1}^{0,L+1}\left[e^y \left| \begin{array}{c} \Xi_L^{(1)}, (1,1,1) \\ \Xi_L^{(1)}, (0,1,1) \end{array} \right. \right], \quad (27)$$

*where the coefficient sets $\Xi_k^{(1)}$ and $\Xi_k^{(2)}$ are defined earlier in (5) and (6) respectively.*

*Proof:* In order to derive the CDF of $Y$, we proceed as follows.

We integrate the PDF expressed in (15) from $0$ through $\gamma$ and obtain the CDF for a single Gamma RV, in terms of Meijer's G function, as

$$P_{\gamma_l}(\gamma_l) = \left(\frac{m_l}{\Omega_l}\right)^{m_l} \mathrm{G}_{m_l+1,m_l+1}^{m_l+1,0}\left[e^{-\gamma_l} \left| \begin{array}{c} \Phi_{m_l}^{(1)}, 1 \\ \Phi_{m_l}^{(2)}, 0 \end{array} \right. \right], \quad (28)$$

where the coefficient sets $\Phi_k^{(1)}$ and $\Phi_k^{(2)}$ are defined earlier in (16) and (17) respectively.

Now, performing a similar integral operation on (4), utilizing a similar explanation as presented in the proof of the PDF of the sum of Gamma RVs i.e. Theorem 1 to obtain (4) from (10), and further making some simple modifications to the Mellin-Barnes integral representation to satisfy the exact definition of the Fox's $\bar{H}$ function, we obtain a final closed-form result for the CDF of $Y$, valid for both integer-order as well as non-integer-order fading parameters, in terms of Fox's $\bar{H}$ function as presented in (27). Hence, in other words, the expression presented in (28) is a special case of (27), Theorem 2 for $L=1$. ∎

As for validation and numerical examples, Fig. 4 and Fig. 5 present the CDF and the logarithmic CDF respectively of the output SNR obtained from the exact closed-form expression (27) and show a perfect match between this obtained closed-form analytical result and the one obtained via Monte Carlo simulations for varying $L's$ (i.e. $L=3,4,5$), and their respective fixed fading parameters $m_1 = 0.6$, $m_2 = 1.1$, $m_3 = 2$, $m_4 = 3.4$, and $m_5 = 4.5$. The logarithmic plots were selected to display the accuracy of the matched results.

*2) Special Case (Integer m Fading Parameters):*

**Corollary 3** (CDF of the Sum of Gamma or Equivalently Squared Nakagami-*m* RVs for Integer-Order Fading Parameters). *In this case using some simple algebraic manipulations, the expression in (27) simplifies to*

$$P_Y(y) = \prod_{l=1}^L \left(\frac{m_l}{\Omega_l}\right)^{m_l} \mathrm{G}_{1+\kappa,1+\kappa}^{1+\kappa,0}\left[e^{-y} \left| \begin{array}{c} \Psi_\kappa^{(1)}, 1 \\ \Psi_\kappa^{(2)}, 0 \end{array} \right. \right], \quad (29)$$

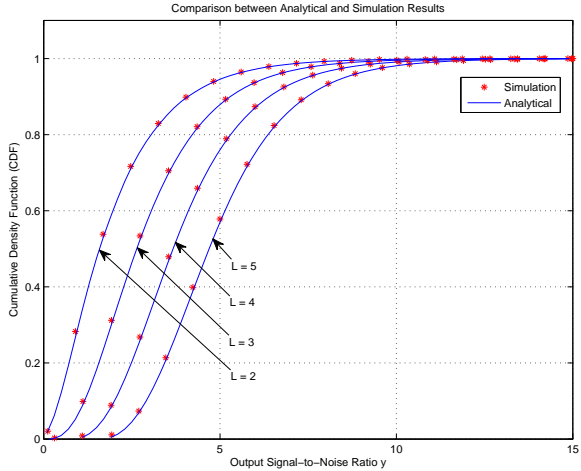

Fig. 4. Comparison between CDFs obtained analytically and via Monte Carlo simulations for varying branches $L$ and respective fixed fading parameters for these channels with $m_1 = 0.6$, $m_2 = 1.1$, $m_3 = 2$, $m_4 = 3.4$, and $m_5 = 4.5$.

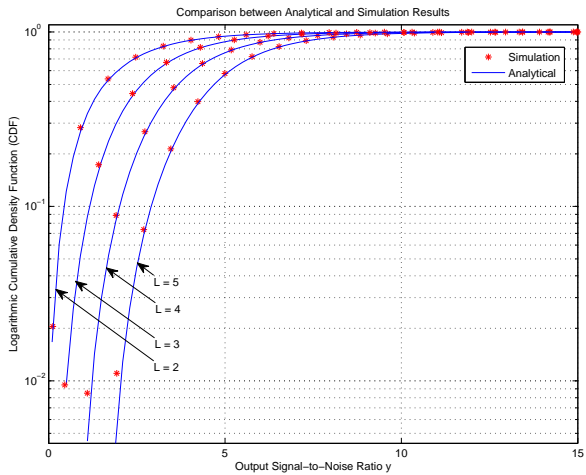

Fig. 5. Comparison between CDFs obtained analytically and via Monte Carlo simulations, on log scale, for varying branches $L$ and respective fixed fading parameters for these channels with $m_1 = 0.6$, $m_2 = 1.1$, $m_3 = 2$, $m_4 = 3.4$, and $m_5 = 4.5$.

*for integer-order fading parameters where the coefficient sets $\Psi_k^{(1)}$ and $\Psi_k^{(2)}$ are defined earlier in (19) and (22) respectively.*

## IV. APPLICATIONS TO THE PERFORMANCE OF DIVERSITY COMBINING RECEIVER SYSTEMS

This section applies the previous results to the performance analysis, in particular outage probability (OP) and BER analysis in Nakagami-$m$ fading environments.

In MRC combining scheme, all the branches are selected at the output. In our case, for a $L$-branch MRC diversity receiver, the signal-to-noise ratio (SNR) $y$, is given by

$$y = \gamma_1 + \cdots + \gamma_L. \quad (30)$$

TABLE III
CONDITIONAL ERROR PROBABILITY (CEP) PARAMETERS

| Modulation | p | q |
|---|---|---|
| Coherent Binary Frequency Shift Keying (CBFSK) | 0.5 | 0.5 |
| Coherent Binary Phase Shift Keying (CBPSK) | 0.5 | 1 |
| Non-Coherent Binary Frequency Shift Keying (NBFSK) | 1 | 0.5 |
| Differential Binary Phase Shift Keying (DBPSK) | 1 | 1 |

### A. Outage Probability

Moreover, when the instantaneous MRC output SNR $y$ falls below a given threshold $y_{\text{th}}$, we encounter a situation labeled as outage and it is an important feature to study outage probability (OP) of a system. Hence, another important fact worth stating here is that the expression derived in the Corollary 3 also serves the purpose for the expression of OP of MRC diversity combining receivers based wireless communication system that is experiencing i.n.i.d. Nakagami-$m$ fading channels or in other words, when the desired user is subject to Nakagami-$m$ fading, the probability that the SNR falls below a predetermined protection ratio $y_{\text{th}}$ can be simply expressed, for both integer-order as well as non-integer-order fading parameters, by replacing $y$ with $y_{\text{th}}$ in (27) as

$$P_{\text{out}}(y_{\text{th}}) = P_Y(y_{\text{th}}). \quad (31)$$

Employing similar substitutions, all the other respective expressions of CDF can be utilized for OP such as replacing $\gamma$ with $\gamma_{\text{th}}$ in (28) and/or replacing $y$ with $y_{\text{th}}$ in (29).

### B. Average BER

The most straightforward approach to obtain BER $P_e$ for MRC is to average the conditional error probability (CEP) $P_e(\epsilon|y)$ for the given SNR given by

$$P_e(\epsilon|y) = \frac{\Gamma(p, qy)}{2\Gamma(p)}, \quad (32)$$

over the PDF of the combiner output SNR [1] i.e.

$$P_e = \int_0^\infty P_e(\epsilon|y) \, p_Y(y) \, dy. \quad (33)$$

The expression in (32) is a unified CEP expression for coherent and non-coherent binary modulation schemes over an AWGN channel [33]. In (32), $\Gamma(\cdot, \cdot)$ is the complementary incomplete Gamma function [24, Eq. (8.350.2)]. The parameters $p$ and $q$ in (32) account for different modulation schemes. For an extensive list of modulation schemes represented by these parameters, one may look into [33], [34] or refer to Table III.

*1) General Case (Non-Integer $m$ Fading Parameters):*

**Theorem 3** (BER of a $L$-branch MRC System Operating over Nakagami-$m$ Fading Channels for Binary Modulation Schemes). *The BER of a $L$-branch MRC diversity combining receiver wireless communication system running over Nakagami-$m$ fading channels, valid for both integer-order as well as non-integer-order fading parameters and for any binary modulation scheme including coherent binary frequency* shift keying (CBFSK), non-coherent binary frequency shift keying (NBFSK), coherent binary phase shift keying (CBPSK), and differential binary phase shift keying (DBPSK), can be expressed in closed-form, in terms of the extended Fox's $\bar{H}$ function $(\hat{H})$[4], as

$$P_e = \frac{q^p}{2} \prod_{l=1}^{L} \left(\frac{m_l}{\Omega_l}\right)^{m_l} \hat{H}_{L+2,L+2}^{L+1,1}\left[1 \left| \begin{array}{c} (1-q,1,p), \zeta_1 \\ \zeta_2, (-q,1,p) \end{array}\right.\right], \quad (34)$$

where

$$\zeta_1 = \Upsilon_L^{(1)}, (1,1,1), \quad (35)$$

and

$$\zeta_2 = (0,1,1), \Upsilon_L^{(2)}, \quad (36)$$

and where the coefficient sets $\Upsilon_k^{(1)}$ and $\Upsilon_k^{(2)}$, $k \in \mathbb{N}$ are defined as

$$\Upsilon_k^{(1)} = \overbrace{\left(1+\frac{m_1}{\Omega_1},1,m_1\right),\ldots,\left(1+\frac{m_k}{\Omega_k},1,m_k\right)}^{k\text{-bracketed terms}}, \quad (37)$$

and

$$\Upsilon_k^{(2)} = \overbrace{\left(\frac{m_1}{\Omega_1},1,m_1\right),\ldots,\left(\frac{m_k}{\Omega_k},1,m_k\right)}^{k\text{-bracketed terms}}, \quad (38)$$

respectively.

*Proof:* Utilizing (33) by substituting (32) and (4) into it and performing some simple manipulations along with some simple rearrangements of $\Gamma(.)$ function terms, we get an exact closed-form result of the integral valid for both integer-order as well as non-integer-order fading parameters and for any binary modulation scheme including CBFSK, NBFSK, CBPSK, and DBPSK, in terms of extended Fox's $\bar{H}$ function $(\hat{H})$, as presented above in (34), Theorem 3. ∎

*2) Special Case (Integer $m$ Fading Parameters):*

**Corollary 4** (BER of a $L$-branch MRC System Operating with Nakagami-$m$ Integer-Order Fading Channels for Coherent BFSK and Coherent BPSK Binary Modulation Schemes)**.** *The above presented BER expression in (34), Theorem 3 is simplified, via simple algebraic manipulations, to the following closed-form expression when considering BFSK and BPSK coherent binary modulation schemes with only integer-order fading parameters. It is represented in terms of the Fox's $\bar{H}$ function as*

$$P_e = \frac{q^p}{2} \prod_{l=1}^{L} \left(\frac{m_l}{\Omega_l}\right)^{m_l} \bar{H}_{\kappa+2,\kappa+2}^{\kappa+1,1}\left[1 \left| \begin{array}{c} (1-q,1,p), \chi_1 \\ \chi_2, (-q,1,p) \end{array}\right.\right], \quad (39)$$

where

$$\chi_1 = \Delta_\kappa^{(1)}, (1,1), \quad (40)$$

and

$$\chi_2 = (0,1), \Delta_\kappa^{(2)}, \quad (41)$$

and where the coefficient sets $\Delta_k^{(1)}$ and $\Delta_k^{(2)}$, $k \in \mathbb{N}$ are defined in (42) and (43) respectively. $K$ is the total number of Gamma or equivalently squared Nakagami-$m$ RVs i.e. $L$

---

[4]The extended Fox's $\bar{H}$ function $(\hat{H})$ was first introduced in [23] and has a MATHEMATICA® implementation given in Table IV

number of total branches for our specific wireless communication system being considered here. It must be noted that this result is numerically equivalent to the result presented in [9, Eq. (18)].

**Corollary 5** (BER of a $L$-branch MRC System Operating over Nakagami-$m$ Integer-Order Fading Channels for Non-Coherent BFSK and Differential BPSK Binary Modulation Schemes)**.** *Further down the line, the BER expression in (39) is simplified, via simple algebraic manipulations, to the following closed-form expression when considering non-coherent BFSK and differential BPSK binary modulation schemes for integer-order only fading parameters. This expression is represented in terms of the Meijer's G function as*

$$P_e = \frac{q^p}{2} \prod_{l=1}^{L} \left(\frac{m_l}{\Omega_l}\right)^{m_l} G_{\kappa+p+1,\kappa+p+1}^{\kappa+1,p}\left[1 \left| \begin{array}{c} o_p^{(1)}, \Psi_\kappa^{(1)}, 1 \\ 0, \Psi_\kappa^{(2)}, o_p^{(2)} \end{array}\right.\right], \quad (44)$$

where the coefficient sets $\kappa$, $\Psi_k^{(1)}$, and $\Psi_k^{(2)}$ are defined earlier in (21), (19), and (22) respectively, the coefficient sets $o_k^{(1)}$ and $o_k^{(2)}$, $k \in \mathbb{N}$ are defined as

$$o_k^{(1)} = \overbrace{(1-q),\ldots,(1-q)}^{k\text{-times}}, \quad (45)$$

and

$$o_k^{(2)} = \overbrace{(-q),\ldots,(-q)}^{k\text{-times}}, \quad (46)$$

respectively. It must be noted that this result is numerically equivalent to the result presented in [9, Eq. (19)].

*3) Numerical Examples and Discussion:* The numerical results for BER of MRC diversity combining receiver scheme with $L$-diversity over i.n.i.d. Gamma or equivalently squared Nakagami-$m$ fading channels are presented in this section.

First we use our MATHEMATICA® implementation of the extended Fox's $\bar{H}$ function $(\hat{H})$ given in Table IV in order to give numerical results based on (34), (39), and/or (44). With this implementation, the extended Fox's $\bar{H}$ function $(\hat{H})$ can be evaluated fast and accurately. This computability, therefore, has been utilized for different digital modulation schemes and is employed to discuss the results in comparison to respective Monte Carlo simulation outcomes.

The average SNR per bit in all the scenarios discussed is assumed to be equal. In addition, different digital modulation schemes are represented based on the values of $p$ and $q$ where $p = 0.5$ and $q = 1$ represents CBPSK, $p = 1$ and $q = 1$ represents DBPSK, CBFSK is represented by $p = 0.5$ and $q = 0.5$, and NBFSK is represented by $p = 1$ and $q = 0.5$. In Monte Carlo simulations, the Gamma or equivalently squared Nakagami-$m$ fading channel generation is readily available in MATLAB.

We observe from Fig. 6 that this implemented computability of extended Fox's $\bar{H}$ function $(\hat{H})$ provides a perfect match to the MATLAB simulated results and the results are as expected i.e. the BER decreases as the SNR increases. Its important to note here that these values for the parameters were selected randomly to prove the validity of the obtained results and

$$\Delta_k^{(1)} = \overbrace{\overbrace{\left(1+\frac{m_1}{\Omega_1},1\right),\ldots,\left(1+\frac{m_1}{\Omega_1},1\right)}^{m_1\text{-times}},\ldots,\overbrace{\left(1+\frac{m_K}{\Omega_K},1\right),\ldots,\left(1+\frac{m_K}{\Omega_K},1\right)}^{m_K\text{-times}}}^{k\text{-bracketed terms}}, \qquad (42)$$

$$\Delta_k^{(2)} = \overbrace{\overbrace{\left(\frac{m_1}{\Omega_1},1\right),\ldots,\left(\frac{m_1}{\Omega_1},1\right)}^{m_1\text{-times}},\ldots,\overbrace{\left(\frac{m_K}{\Omega_K},1\right),\ldots,\left(\frac{m_K}{\Omega_K},1\right)}^{m_K\text{-times}}}^{k\text{-bracketed terms}} \qquad (43)$$

TABLE IV
MATHEMATICA® IMPLEMENTATION OF THE EXTENDED FOX'S $\bar{\text{H}}$ FUNCTION ($\hat{H}$)

```
(*Extended Fox H-Bar-Function*)
Clear[x, Ω];
ExtendedFoxHBar::InconsistentCoeffs = "Inconsistent coefficients!";
ExtendedFoxHBar[a_, b_, z_] := Module[
  {Z, s, Pa, Pb, Qa, Qb, M, R, Rmax, Rmin, value},
  (*Gamma product terms*)
  Pa = Function[u, Product[
    Power[Gamma[1 - a[[1, n, 1]] - u a[[1, n, 2]]], a[[1, n, 3]]], {n, 1, Length[a[[1]]]}]];
  Qa = Function[u, Product[Power[Gamma[a[[2, n, 1]] + u a[[2, n, 2]]], a[[2, n, 3]]],
    {n, 1, Length[a[[2]]]}]];
  Pb = Function[u, Product[Power[Gamma[b[[1, n, 1]] + u b[[1, n, 2]]], b[[1, n, 3]]],
    {n, 1, Length[b[[1]]]}]];
  Qb = Function[u, Product[Power[Gamma[1 - b[[2, n, 1]] - u b[[2, n, 2]]], b[[2, n, 3]]],
    {n, 1, Length[b[[2]]]}]];
  M = Function[u, Pa[u] Pb[u] / Qa[u] / Qb[u]];
  (*Contour Limiter*)
  (*Depends on numerator argument i.e. it
   must be at least half of the least valued gamma arguments*)
  {Rmin, Rmax} = {Max[-Min[b[[1, All, 1]] / b[[1, All, 2]]], -Infinity],
    Min[Min[(1 - a[[1, All, 1]]) / a[[1, All, 2]]], Infinity]};
  If[Rmin == -Infinity && Rmax ≠ Infinity, Rmin = Rmax - 1];
  If[Rmin ≠ -Infinity && Rmax == Infinity, Rmax = Rmin + 0.1];
  If[Rmin == Rmax, Message[ExtendedFoxHBar::InconsistentCoeffs]];
  R = Mean[{Rmax, Rmin}];
  (*Assignment and Declaration*)
  Z = z;
  (*Final Evaluation*)
  value =
    1
    ─── NIntegrate[M[s] Z^-s, {s, -50 - I 50, R - I 50, R + I 50, -50 + I 50}, MaxRecursion → 55];
   2 π I
  (*Returning back the value*)
  Return[value];
];
(*End of ExtendedFoxHBar*)
```

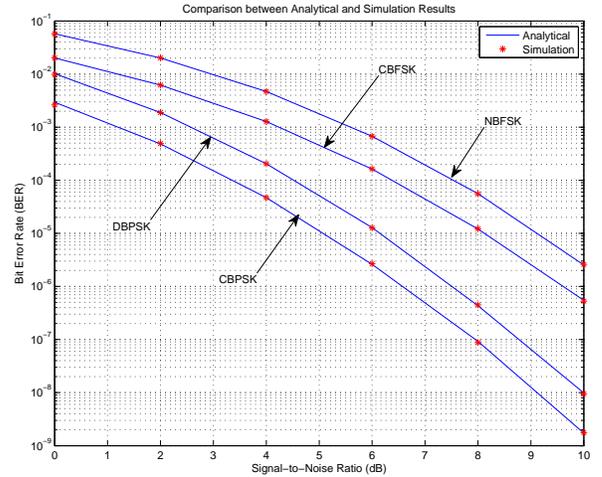

Fig. 6. Average BER of different binary modulation schemes over i.n.i.d. Gamma or equivalently squared Nakagami-$m$ fading channels with $L = 5$-branch MRC and fading parameters for these channels with $m_1 = 0.6$, $m_2 = 1.1$, $m_3 = 2$, $m_4 = 3.4$, and $m_5 = 4.5$.

hence specific values based on the standards can be used to obtain the required results.

Furthermore, it can be seen from Fig. 6 that, as expected, CBPSK outperforms the other modulation schemes and the coherent binary modulation schemes outperform their respective non-coherent and/or differential binary modulation scheme i.e. CBPSK outperforms DBPSK and CBFSK outperforms NBFSK. Additionally, PSK in general performs better than FSK, as expected. Similar results for any other values of $m's$ can be observed for the exact closed-form BER for $L$-diversity i.n.i.d. Gamma or equivalently squared Nakagami-$m$ channels presented in this work.

## V. CONCLUDING REMARKS

We derived alternative closed-form expressions for the PDF and the CDF of the sum of i.n.i.d. Gamma or equivalently squared Nakagami-$m$ RVs in the case of both integer-order as well as non-integer-order fading figure parameters. An interesting finding is that these expressions can be written in terms of special functions, specifically Fox's $\bar{\text{H}}$ functions. Based on these statistical formulas obtained, and following a PDF-based approach, we analyzed the performance of a MRC diversity combining receiver based wireless communication system operating over i.n.i.d. Nakagami-fading channels and important performance metrics such as OP and BER were expressed in closed form and hence this serves as the key feature along side the novel statistical derivations of PDF and CDF. For instance, an exact closed-form expression for the BER performance of different binary modulations with $L$-branch MRC scheme over i.n.i.d. Gamma or equivalently squared Nakagami-$m$ fading channels was derived. The analytical calculations were done utilizing a general class of special functions including Meijer's G function, Fox's $\bar{\text{H}}$ function, and extended Fox's $\bar{\text{H}}$ function ($\hat{H}$). Our results complement previously published results that are either in the form of infinite sums or nested sums or recursive expressions or higher order derivatives of the fading parameter. In addition, this work presents numerical examples to validate and illustrate the mathematical formulation developed in this work and to show the effect of the fading severity and unbalance on the system performance.

## ACKNOWLEDGMENT

We would like to thank King Abdullah University of Science and Technology (KAUST), Thuwal, Makkah Province, Saudi Arabia for providing support and resources respectively for this research work.